\documentclass[lettersize,journal]{IEEEtran}
\usepackage{amsmath,amsfonts, amssymb}
\usepackage{algorithmic}
\usepackage{algorithm}
\usepackage{array}
\usepackage[caption=false,font=normalsize,labelfont=sf,textfont=sf]{subfig}
\usepackage{textcomp}
\usepackage{stfloats}
\usepackage{url}
\usepackage{verbatim}
\usepackage{graphicx}
\usepackage{booktabs}
\usepackage{cite}
\usepackage[hidelinks]{hyperref}
\usepackage{cleveref}
\usepackage{siunitx}

\newtheorem{theorem}{Theorem}
\newtheorem{corollary}{Corollary}
\newtheorem{proposition}{Proposition}
\newtheorem{lemma}{Lemma}

\begin{document}

\title{Explicit Convex Inner Approximation of TSO-DSO Flexibility Regions via Exact Second-order Cone Relaxation in Radial Networks}

\author{Wenyu Liu,~\IEEEmembership{Graduate Student Member,~IEEE} and Mario Paolone,~\IEEEmembership{Fellow,~IEEE}
\thanks{The authors are with the Distributed Electrical Systems Laboratory, Swiss Federal Institute of Technology of Lausanne, 1015 Lausanne, Switzerland, email: \{wenyu.liu, mario.paolone\}@epfl.ch.}
}

\markboth{}%
{Liu \MakeLowercase{\textit{et al.}}: A Sample Article Using IEEEtran.cls for IEEE Journals}

\IEEEpubid{}

\maketitle

\begin{abstract}
Accurate representations of the flexibility in distribution grids via tractable algorithms are essential for TSO–DSO coordination. However, flexibility region (FR) estimation based on nonlinear optimal power flow (OPF) generally lacks feasibility guarantees, while conventional convex relaxations may suffer from inexactness. This paper proposes a framework for constructing explicit inner approximations of active–reactive power FRs in radial networks based on the augmented relaxed OPF (AR-OPF). First, less-conservative sufficient conditions for the AR-OPF exactness are derived and extended to generalized objectives. Furthermore, an interior-feasibility analysis is established, guaranteeing that the entire region enclosed by the traced boundary is AC feasible under the proposed sufficient conditions. Building on these theoretical results, two adaptive algorithms are developed to construct explicit inner approximations and numerically convex representations of FRs. Numerical evaluations on different benchmark networks demonstrate consistently small relaxation residuals and high flexibility area retention in most cases, showing that the proposed methods provide a grounded approach to construct operationally admissible FRs for TSO–DSO coordination.
\end{abstract}


\section{Introduction}

Increasing integration of distributed energy resources (DERs), energy storage systems, and flexible loads has transformed traditionally inelastic distribution grids into active assets contributing to the power system operation. By coordinating these resources, distribution system operators (DSOs) can provide active  and reactive power flexibility to transmission system operators (TSOs), supporting congestion management, voltage and frequency regulation, and the efficient utilization of existing network infrastructure \cite{usmanNovelTwostageTSO2023,valverdeCoordinationDistributedReactive2019,jacobsProbabilisticFlexibilityAggregation2026,liNetloadRangeCost2026}. However, exploiting such flexibility at the transmission level requires an accurate characterization of the aggregate capabilities of the underlying distribution network, without exposing its detailed topology, operating constraints, or individual resource characteristics \cite{valverdeTSODSOCoordinationFlexibility2027}. To this end, early real-time control frameworks such as Commelec \cite{bernsteinComposableMethodRealTime2015} abstracted local resource operational capabilities into dynamic active-reactive power profiles. Subsequent researches proposed the so-called flexibility region (FR), conceptually tied to Commelec, provide an interface-level representation of the admissible active and reactive power exchanges between transmission and distribution networks, subject to the physical and operational constraints of the distribution system \cite{silvaEstimatingActiveReactive2018}. Such a representation allows distribution-level flexibility to be incorporated into transmission-level optimization and market-clearing frameworks in a tractable manner. Importantly, the reliability of the FR is essential: an overestimated region may lead to infeasible flexibility requests, whereas an excessively conservative approximation unnecessarily restricts the available flexibility \cite{barzegarDynamicOperatingEnvelopes2026}. Therefore, constructing accurate and operationally admissible FRs with computationally tractable algorithms is a fundamental requirement for effective TSO-DSO coordination.

\subsection{Flexibility Aggregation Problems}

Estimating the FR generally involves solving power flow constrained optimization problems to determine the achievable interface power exchanges under voltage, branch-flow, and distributed resource operational limits. Early approaches employed nonlinear AC power flow models and Monte Carlo simulations to estimate the available flexibility \cite{helenoEstimationFlexibilityRange2015}. Subsequently, directional boundary tracing methods were developed in \cite{silvaEstimatingActiveReactive2018,capitanescuTSODSOInteraction2018}, respectively, providing more structured and computationally tractable approximations. However, these methods rely on nonconvex optimization and primarily identify boundary points, without establishing the feasibility of the entire enclosed region. For example, a test case in \cite{prionistisFlexibilityRegionDiscontinuity2025} shows that the enclosed region may be discontinuous and exist unreachable holes. To improve tractability, linearized power flow formulations have been employed, particularly for multi-timestep flexibility aggregation \cite{chenAggregatePowerFlexibility2020,cuiNetworkCognizantTimeCoupledAggregate2021}. Nevertheless, deviations from the nonlinear AC equations may result in inaccurate representations of network feasibility. Alternatively, second-order cone programming (SOCP) relaxations have been investigated for FR estimation \cite{prionistisFastMethodApproximate2023,moringExploringInexactnessSecondOrder2026}. Although these approaches exploit the computation efficiency of the convex optimization, inexactness of the relaxed branch-flow constraints may result-in physically impossible points that violates the original AC power flow equations. Furthermore, recent studies have demonstrated that FRs can exhibit non-convexity and disconnected feasible components \cite{pozoTSODSOFlexibilityRegions2025,prionistisFlexibilityRegionDiscontinuity2025}. Consequently, neither the convex hull of sampled boundary points nor the region enclosed by a numerically traced boundary is necessarily AC feasible. This is particularly relevant when the FR is transferred to the TSO as an explicit feasible set for higher-level optimization. These observations motivate a framework that combines verifiable exactness of convex relaxations with theoretical guarantees on interior feasibility and practical construction of convex inner approximations.

\subsection{Contribution of the Paper}

This paper addresses these challenges by extending the augmented relaxed optimal power flow (AR-OPF) formulation of \cite{nickExactConvexFormulation2018} to explicit FR assessment. The main contributions are summarized as follows:

\begin{enumerate}
    \item \textit{Improved exactness conditions:} Less-conservative sufficient conditions for AR-OPF exactness are derived, extending \cite{nickExactConvexFormulation2018}. The analysis is further generalized to objectives for losses-regularized boundary tracing.

    \item \textit{Interior feasibility guarantee:} We establish conditions under which the entire region enclosed by a continuously traced AR-OPF boundary is AC feasible. By exploiting the monotonicity of the exactness certificate with respect to losses regularization and the contraction of the traced boundary, this result extends exactness guarantees from boundary points to an explicit inner FR.

    \item \textit{Adaptive explicit FR construction:} Two adaptive algorithms are developed to obtain explicit, numerically validated inner approximations and convex inner representations of FRs. Their accuracy, conservativeness, and computational scalability are evaluated using radial networks of different sizes and configurations.
\end{enumerate}

\section{Problem Statement}

\subsection{Notations and Definitions}
\begin{figure}[ht]
    \centering
    \includegraphics[width=.75\linewidth]{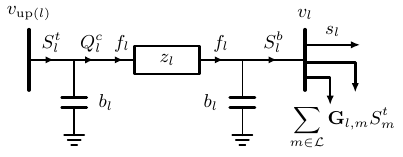}
    \caption{Classic two-port $\Pi$ model of a passive reciprocal branch adopted for the formulation of the OPF relaxed constraints\cite{nickExactConvexFormulation2018}.}
    \label{fig-nick-PI}
\end{figure}
We follow the notation and formulation of \cite{nickExactConvexFormulation2018}. Hereafter, the prefix ``R.'' identifies equations and results of \cite{nickExactConvexFormulation2018} (e.g., (R.1) and \textit{Lemma R.I} denote Eq. (1) and \textit{Lemma I} therein), whereas unprefixed labels refer to the present paper. Symbols and matrices inherited from \cite{nickExactConvexFormulation2018} retain their original definitions unless otherwise stated. Specifically, the network topology is radial; index 0 represents the slack bus that has a fixed voltage magnitude $v_0$; only bus 1 is connected to the slack bus, otherwise the problem is applied separately to each radial network connected to the slack bus; $\mathcal L = \{1,2,\dots, L\}$ denotes the set of non-slack buses, $\operatorname{up}(l)$ is the label of the up-streaming bus of bus $l$, and, equivalently, the set of distribution lines indexed by their downstream buses; $\mathbf I$ denotes the identity matrix, $\mathbf G$ is the oriented adjacency matrix, i.e. $\mathbf G_{i,k} =1$ if $k = \operatorname{up}(l)$ and 0 otherwise, and $\mathbf H=(\mathbf I-\mathbf G)^{-1}$ is the transitive closure of $\mathbf G$. Vectors $r$, $x$, and $2b$ collect the line resistance, reactance, and shunt-susceptance parameters, respectively. Notation without subscript denote column vectors with $L$ rows, such as $v=[v_1, \dots, v_L]^\top$. We assume the network is composed of passive, reciprocal two-port branches, as shown in \Cref{fig-nick-PI}. This model is used in the power flow calculation, in which the variables, $s$, $P^t$, $Q^t$, $Q^c$, $v$, and $f$ denote, respectively, the power absorption, the branch upper side active power, branch upper side reactive power, central-element reactive power, squared bus-voltage magnitude, and squared current flow in the central part of the two-port $\Pi$ branch model.

Let $y = [P_1^t, Q_1^t]^\top\in\mathbb{R}^2$ denote the active and reactive power injected from the TSO to the DSO network. The exact FR at the TSO-DSO interface is defined as
$$
\mathcal Y_{\rm AC}
=
\left\{
y\in\mathbb R^2
\;\middle|\;
\begin{array}{l}
\exists\,(s,v)
\text{ s.t.}\\
\mathfrak f(s,v,y)= 0,\ \mathfrak g(s,v,y)\le 0
\end{array}
\right\},
$$
where $s$ collects the controllable and uncontrollable nodal absorptions at non-slack buses, $\mathfrak f(\cdot)$ represents the AC power flow equations, and $\mathfrak g(\cdot)$ collects the network and resource constraints. According to the non-approximated branch flow model, the power flow equations $\mathfrak f(\cdot)$ are given by \eqref{eq-AC-pf}, $\forall\, l \in \mathcal{L}$:
\begin{subequations}\label{eq-AC-pf}
\begin{align}
S_l^t &= s_l + \sum_{m \in \mathcal{L}} \mathbf{G}_{l,m} \, S_m^t + z_l f_l - j \left( v_{\operatorname{up}(l)} + v_l \right) b_l\label{eq-AC-pf-St}\\
v_l &= v_{\operatorname{up}(l)} - 2 \Re \left( z_l^* \left( S_l^t + j v_{\operatorname{up}(l)} b_l \right) \right) + |z_l|^2 f_l \label{eq-AC-pf-v}\\
f_l &= \frac{\left| S_l^t + j v_{\operatorname{up}(l)} b_l \right|^2}{v_{\operatorname{up}(l)}} = \frac{\left| S_l^b - j v_l b_l \right|^2}{v_l} \label{eq-AC-pf-f}\\
S_l^b &= s_l + \sum_{m \in \mathcal{L}} \mathbf{G}_{l,m} \, S_m^t, \label{eq-AC-pf-Sb}
\end{align}
\end{subequations}
where $j:=\sqrt{-1}$, $\mathfrak R(\cdot)$ and $\mathfrak I(\cdot)$ denotes the real and imaginary parts of complex numbers. Without losing generality, $\mathfrak g(\cdot)$ can be given by \eqref{eq-g-cons}, $\forall\, l \in \mathcal{L}$:
\begin{subequations}\label{eq-g-cons}
\begin{align}
& v^{\min} \le v_l \le v^{\max}\\
& \left| S_l^b \right|^2 \le I_l^{\max} v_l\\
& \left| S_l^t \right|^2 \le I_l^{\max} v_{\operatorname{up}(l)}\\
& p_l^{\min} \le \Re\left(s_l\right) \le p_l^{\max} \label{eq-g-cons-res-P}\\
& q_l^{\min} \le \Im\left(s_l\right) \le q_l^{\max}, \label{eq-g-cons-res-Q}
\end{align}
\end{subequations}
where $v^{\min}$ and $v^{\max}$ are the square of minimum and maximum magnitude of nodal voltages, $I^{\max}$ is the square of branch current limits, and the resource absorption limits are given by \eqref{eq-g-cons-res-P} and \eqref{eq-g-cons-res-Q}. Note that $I^{\max}$ applies to the branch terminal currents, not to $f$.

The difficulty is that $\mathcal Y$ is defined implicitly through a nonlinear projection. Owing to the non-convex AC power flow equations, the resulting region is not guaranteed to be convex and, under stressed operating conditions, may exhibit pronounced non-convexities \cite{pozoTSODSOFlexibilityRegions2025} or disconnected feasible portions \cite{prionistisFlexibilityRegionDiscontinuity2025}. Direct construction of the complete set is therefore generally intractable for networks of large size.

\subsection{Directional Boundary Tracing with Non-linear Power Flow}

A common approach is to characterize the FR explicitly through a finite number of directional optimization problems \cite{silvaEstimatingActiveReactive2018}. Let $d_\gamma = [\cos\gamma, \sin\gamma]^\top, \gamma\in[0,2\pi)$ denote a search direction in the slack-power plane. A boundary point of the AC FR can be obtained by solving
\begin{equation}\label{eq-opt-silva}
\begin{aligned}
\min_{s, S, v, f}
\quad &
d_\gamma^\top y = \cos\gamma \cdot P_1^t + \sin\gamma \cdot Q_1^t
\\
\text{s.t.}\quad & \eqref{eq-AC-pf} \ \text{and}\ \eqref{eq-g-cons}.
\end{aligned}
\end{equation}
Repeating \eqref{eq-opt-silva} over a set of directions $\Gamma = \{\gamma_1,\ldots,\gamma_N\}$ generates the boundary samples $\mathcal S_{\rm AC} = \left\{ y_{\rm AC}^\star(\gamma) : \gamma\in\Gamma \right\}$. When \eqref{eq-opt-silva} is solved globally, each point in $\mathcal S_{\rm AC}$ is physically feasible and extremal in the corresponding direction. In practice, however, \eqref{eq-opt-silva} is a non-convex AC optimal power flow problem. Repeatedly solving it over a dense angular discretization may therefore be computationally demanding and can be sensitive to initialization and local optima. Furthermore, as the network size increases, the computation time for nonlinear programming may increase significantly.

\subsection{Opportunities and Challenges of SOCP Relaxations}
Convex relaxations of the AC power flow equations offer an attractive alternative. For radial networks, second-order cone programming (SOCP) relaxations are particularly appealing because the resulting boundary tracing problems can be solved efficiently and reliably \cite{lowConvexRelaxationOptimal2014a}. We observe that \eqref{eq-opt-silva} is non-convex due to \eqref{eq-AC-pf-f}, it becomes convex if we apply the classical SOCP relaxation in \eqref{eq-AC-pf-f-soc}:
\begin{equation}
    f_l \ge \frac{\left| S_l^t + j v_{\operatorname{up}(l)} b_l \right|^2}{v_{\operatorname{up}(l)}}, \quad \forall l\in\mathcal{L} \label{eq-AC-pf-f-soc}
\end{equation}
The relaxation is exact at a solution if all equalities in \eqref{eq-AC-pf-f-soc} are binding. In that case, the relaxed problem admits a physically valid AC realization. Otherwise, the resulting interface point may belong to the relaxed solution while being infeasible for the real AC network.

Existing SOCP-based FR approaches, e.g. \cite{prionistisFastMethodApproximate2023, moringExploringInexactnessSecondOrder2026} therefore face the exactness challenges: the exactness generally has to be assessed after optimization or is guaranteed only under very restrictive conditions (e.g., there is no bound on nodal reactive power \cite{lowConvexRelaxationOptimal2014a}) that are hard to fulfill for real systems. To address this issue, this work adopts the augmented relaxed optimal power flow (AR-OPF) formulation proposed by \cite{nickExactConvexFormulation2018}, in which additional convex constraints are introduced to augment the original SOCP relaxation. The AR-OPF takes the following form:
\begin{equation}\label{eq-opt-AR-OPF}
\begin{aligned}
\min_{s, S, v, f, \hat{S}, \bar{v}, \bar{S}, \bar{f}}
\quad &
J = \sum_{l \in \mathcal{L}}\left[\mathcal{C}\left(\mathfrak{R}(s_l), \mathfrak{I}(s_l) \right) \right] + \mathcal{C}^e\left(P_1^t\right),
\\
\text{s.t.}\quad & \eqref{eq-AC-pf-St},  \eqref{eq-AC-pf-v},  \eqref{eq-AC-pf-Sb}, \eqref{eq-g-cons-res-P}, \eqref{eq-g-cons-res-Q}, \eqref{eq-AC-pf-f-soc}, \ \text{and}\\
&\text{augmented con. (R.11), (R.12.c-R.12.g)}.
\end{aligned}
\end{equation}
where $\mathcal C(\cdot)$ is the cost function of nodal absorption, and $\mathcal C^e(P_1^t)$ is strictly increasing in the active power imported from the external grid. Both $\mathcal C(\cdot)$ and $\mathcal C^e(P_1^t)$ are assumed to be convex. Moreover, a set of sufficient conditions for the exactness of the SOCP relaxation are also provided (R.C1-R.C5).

The usefulness of AR-OPF for approximating FRs, however, depends critically on its exactness properties. In particular, two issues must be addressed before it can be employed for boundary tracing:
\begin{itemize}
    \item First, the sufficient exactness conditions originally associated by R.C1--R.C5 can become restrictive for large networks. Although the AR-OPF almost always remain numerically exact in practice, the available sufficient conditions cannot certify it \cite{boboSecondorderConeRelaxations2021}. Less conservative conditions are therefore desirable to enlarge the range of operating conditions for which exactness can be verified ex ante.
    \item Second, boundary tracing requires an objective function that varies both $P_1^t$ and $Q_1^t$, The inclusion of $Q_1^t$ is nontrivial from the standpoint of existing AR-OPF exactness arguments, which rely on specific monotonicity properties of the objective. A rigorous characterization of the admissible objective therefore constitutes a prerequisite for applying AR-OPF to boundary tracing.
\end{itemize}
The two issues are tackled below in \Cref{sec-less-conservative-conditions} and \Cref{sec-slack-Q-in-obj} respectively.

\section{AR-OPF Exactness Conditions for Flexibility Region Approximation}

We retain the auxiliary matrices $\mathbf C$, $\mathbf D$, $\mathbf E$, and $\mathbf F$ defined in \cite{nickExactConvexFormulation2018}, where $\mathbf D,\mathbf E,\mathbf F$ are entrywise nonnegative under the conditions considered below. Superscripts $(n)$ denote the iterates of the power flow recovery procedure of \textit{Algorithm R.1}. We then introduce the following new notations:
$$
\mathbf A:=\mathbf H\operatorname{diag}(r),
\ 
e_1:=\begin{bmatrix}
1 & 0 & \cdots & 0
\end{bmatrix}^{\top}\in\mathbb R^L,
\ 
\mathbf a_1:= e_1\mathbf A,
$$
$$
\mathbf a_P^\top:=e_1^\top\mathbf A,
\quad
\mathbf a_Q^\top:=e_1^\top\mathbf F,
\quad
\mathbf R_E := \sum_{k=1}^{\infty}\mathbf E^k = \mathbf E(\mathbf I-\mathbf E)^{-1}.
$$

\subsection{Less Conservative Sufficient Conditions for AR-OPF Exactness}\label{sec-less-conservative-conditions}

The original conditions R.C1--R.C5 guarantee exactness by imposing geometric contraction bounds on the iterative power flow construction of \textit{Algorithm R.1}. These conditions are stronger than required by the proof of \textit{Theorem R.I}. In particular, feasibility recovery requires only convergence of the iteration and preservation of the voltage limits, whereas exactness at an optimum requires the cumulative corrections not to offset the initial reduction of the slack active power. We therefore replace the one-step geometric bounds of R.C1--R.C5 by conditions directly controlling these cumulative effects. The treatment of the branch-ampacity constraints remains unchanged and follows from the augmented variables and \textit{Lemmas R.I--R.II}. Original conditions R.C1--R.C5 collapse into following \eqref{eq-C1-prime} to \eqref{eq-C4-prime}:
\begin{align}
    &\rho(\mathbf H^\top\mathbf M)<1; \tag{C1'}\label{eq-C1-prime} \\
    &  \rho(\mathbf E)<1; \tag{C2'}\label{eq-C2-prime}\\
    & \mathbf D\mathbf R_E\le\mathbf D; \tag{C3'}\label{eq-C3-prime}\\
    & \mathbf a_1^\top\mathbf R_E < \tag{C4'}\mathbf a_1^\top. \label{eq-C4-prime}
\end{align}
$\rho(\cdot)$ denotes the spectral radius of a square matrix and all vector and matrix inequalities are component-wise.

We first show that \eqref{eq-C1-prime} provides the same structural property as \textit{R.C1} while being less conservative. Recall that $\mathbf H^\top = (\mathbf I - \mathbf G^\top)^{-1} \ge 0$ due to the acyclic structure of the radial network. Factoring out $(\mathbf I - \mathbf G^\top)$ yields
\begin{equation}
\mathbf I - \mathbf G^\top - \mathbf M = (\mathbf I - \mathbf G^\top)(\mathbf I - \mathbf H^\top \mathbf M).
\end{equation}
Since $\mathbf H^\top \mathbf M \ge 0$ ($\mathbf M \ge 0$ by definition), \eqref{eq-C1-prime} ensures that $(\mathbf I - \mathbf H^\top \mathbf M)$ is invertible with a convergent, entrywise non-negative Neumann series:
\begin{equation}
(\mathbf I - \mathbf H^\top \mathbf M)^{-1} = \sum_{k=0}^{\infty} (\mathbf H^\top \mathbf M)^k \ge 0.
\end{equation}
Applying the inverse of a matrix product, we have
\begin{equation}
\mathbf C := (\mathbf I - \mathbf G^\top - \mathbf M)^{-1} = (\mathbf I - \mathbf H^\top \mathbf M)^{-1} \mathbf H^\top \ge 0.
\label{eq-C-nonnegative}
\end{equation}
Hence, the existence and entrywise non-negativity of $\mathbf C$ (and consequently $\mathbf D$ and $\mathbf F$) are guaranteed. Finally, because $\rho(\mathbf H^\top \mathbf M) \le \|\mathbf H^\top \mathbf M\|$ for any induced norm, \textit{R.C1} implies \eqref{eq-C1-prime}, but not conversely.

For a particular feasible AR-OPF point, \eqref{eq-C3-prime} and \eqref{eq-C4-prime} can be further relaxed. Let $d:=f^{(0)}-f^{(1)}\ge0$. The corresponding solution-dependent conditions are
\begin{align}
    \mathbf D\mathbf R_Ed
    &\le
    \mathbf Dd,
    \tag{C3'-local}
    \label{eq-C3-prime-local}
    \\
    \mathbf a_1^\top\mathbf R_Ed
    &<
    \mathbf a_1^\top d,
    \qquad d\neq0.
    \tag{C4'-local}
    \label{eq-C4-prime-local}
\end{align}
\eqref{eq-C3-prime} and \eqref{eq-C4-prime} imply their local counterparts for every $d\ge0$.

\begin{theorem}[Revised \textit{Theorem R.I}]
\label{thm-revised-exactness}
\begin{enumerate}
    \item Under \eqref{eq-C1-prime}--\eqref{eq-C3-prime}, for every feasible AR-OPF solution there exists a recovered solution with the same $s$ that satisfies the AC power flow equations, voltage limits, and branch-ampacity constraints, and is therefore feasible for A-OPF and O-OPF.
    \item If \eqref{eq-C4-prime} additionally holds, every optimal solution of AR-OPF satisfies \eqref{eq-AC-pf-f} on every branch and is therefore an optimal solution of A-OPF and O-OPF.
\end{enumerate}
\end{theorem}

The proof relies on the following revision of \textit{Lemma R.III}.

\begin{lemma}[Revised \textit{Lemma R.III}]
\label{lem-revised-RIII}
Let $\omega^{\mathrm{AR}} = (s, S^{(0)}, v^{(0)},\allowbreak f^{(0)}, \hat{S},\allowbreak \bar{v}, \bar{f}, \bar{S})$ be a feasible solution of AR-OPF, and let
$(P^{(n)},Q^{c(n)},v^{(n)},f^{(n)})$
be the sequence generated by \textit{Algorithm R.1}. Define $\Delta f^{(n)}:=f^{(n)}-f^{(n-1)}$, then $d=f^{(0)}-f^{(1)}\ge0$, and $d\neq0$ whenever the relaxed \eqref{eq-AC-pf-f-soc} inequality is strict for at least one line.

Under \eqref{eq-C1-prime}--\eqref{eq-C2-prime},
\begin{equation}
    |\Delta f^{(n)}|
    \le
    \mathbf E^{n-1}d,
    \qquad n\ge1,
    \label{eq-delta-f-new}
\end{equation}
and the sequence generated by \textit{Algorithm R.1} converges.\\
If \eqref{eq-C3-prime-local} additionally holds,
\begin{equation}
    v^{(n)}
    \ge
    v^{(0)},
    \qquad n\ge1.
    \label{eq-v-new}
\end{equation}
If \eqref{eq-C4-prime-local} also holds and $d\neq0$,
\begin{equation}
    P_1^{t(n)}
    <
    P_1^{t(0)},
    \qquad n\ge1.
    \label{eq-P1-new}
\end{equation}
\end{lemma}

\begin{IEEEproof}
The Jacobian bound used in the proof of \textit{Lemma R.III}, i.e., (R.62) gives $|\Delta f^{(n+1)}|\le\mathbf E|\Delta f^{(n)}|$. Since $|\Delta f^{(1)}|=d$, induction yields \eqref{eq-delta-f-new}. This step does not require a norm bound on $\mathbf E$.

Under \eqref{eq-C2-prime}, $\rho(\mathbf E)<1$, and hence $\sum_{k=0}^{\infty}\mathbf E^k=(\mathbf I-\mathbf E)^{-1}$. Consequently,
\begin{equation}
    \sum_{n=1}^{\infty}
    |\Delta f^{(n)}|
    \le
    (\mathbf I-\mathbf E)^{-1}d,
\end{equation}
so $f^{(n)}$ is componentwise Cauchy and converges to some $f^*$. From (R.16), (R.18), and (R.20), we have
\begin{equation}
\begin{aligned}
    P^{(n)}=\hat P+&\mathbf A f^{(n)}, \qquad
    Q^{c(n)}=\hat Q^c+\mathbf F f^{(n)}, \\
    &v^{(n)}=\bar v-\mathbf D f^{(n)}
    \end{aligned}\label{eq-P-Q-v-f-iteration}
\end{equation}
$P^{(n)}$, $Q^{c(n)}$, and $v^{(n)}$ converge as well. From \eqref{eq-P-Q-v-f-iteration},
$$
\begin{aligned}
v^{(n)}-v^{(0)}
&=
-\mathbf D
\sum_{k=1}^{n}\Delta f^{(k)}
 =
\mathbf Dd
-
\mathbf D
\sum_{k=2}^{n}\Delta f^{(k)}\\
&\ge
\mathbf Dd
-
\mathbf D
\sum_{j=1}^{n-1}\mathbf E^j d
\ge
\mathbf Dd-\mathbf D\mathbf R_Ed.
\end{aligned}
$$
Hence \eqref{eq-C3-prime-local} implies \eqref{eq-v-new}. Likewise, using \eqref{eq-P-Q-v-f-iteration},
$$\begin{aligned}
P_1^{t(n)}-P_1^{t(0)}
&=
\mathbf a_1^\top
\sum_{k=1}^{n}\Delta f^{(k)}
=
-\mathbf a_1^\top d
+
\mathbf a_1^\top
\sum_{k=2}^{n}\Delta f^{(k)}
\\
&\le
-\mathbf a_1^\top d
+
\mathbf a_1^\top\mathbf R_Ed.
\label{eq-P-tail}
\end{aligned}$$
Therefore \eqref{eq-C4-prime-local} gives
$P_1^{t(n)}<P_1^{t(0)}$ for all $n\ge1$, and, by taking the limit, $P_1^{t*}<P_1^{t(0)}$.
\end{IEEEproof}

\begin{IEEEproof}[Proof of Theorem~\ref{thm-revised-exactness}, Item~1)]
Let $\omega^{\mathrm{AR}} = (s, S^{(0)}, v^{(0)},\allowbreak f^{(0)}, \hat{S},\allowbreak \bar{v}, \bar{f}, \bar{S})$ be an arbitrary feasible AR-OPF solution. By
Lemma~\ref{lem-revised-RIII},
$$
(P^{(n)},Q^{c(n)},v^{(n)},f^{(n)})
\longrightarrow
(P^*,Q^{c*},v^*,f^*).
$$
Step~1 of \textit{Algorithm R.1} imposes
\begin{equation}
f_l^{(n)}
=
\frac{
\left(P_l^{t(n-1)}\right)^2
+
\left(Q_l^{c(n-1)}\right)^2
}{
v_{\operatorname{up}(l)}^{(n-1)}
},
\qquad
l\in\mathcal L.
\end{equation}
Taking the limit gives
\begin{equation}
f_l^*
=
\frac{
(P_l^{t*})^2+(Q_l^{c*})^2
}{
v_{\operatorname{up}(l)}^*
},
\qquad
\forall l\in\mathcal L.
\label{eq-f-convergence}
\end{equation}
Thus the original \eqref{eq-AC-pf-f} is recovered on every line. Furthermore, \eqref{eq-v-new} and feasibility of the initial AR-OPF point yield $v^*\ge v^{(0)}\ge v^{\min}$.
Since $f^{(n)}\ge0$ by construction and $\mathbf D\ge0$,
\begin{equation}
    v^{(n)}
    =
    \bar v-\mathbf Df^{(n)}
    \le
    \bar v
    \le
    v^{\max},
\end{equation}
and hence
\begin{equation}
    v^{\min}
    \le
    v^*
    \le
    \bar v
    \le
    v^{\max}.
\label{eq-v-limit-bounds}
\end{equation}

Equation~\eqref{eq-f-convergence} establishes the physical branch-current relation \eqref{eq-AC-pf-f}; the terminal ampacity constraints follow separately as in the original proof of \textit{Theorem R.I}. Specifically, \textit{Lemma R.I} constructs the corresponding augmented flow bounds, whose AR-OPF security constraints imply the original terminal-current limits, while \textit{Lemma R.II} establishes O-OPF feasibility. Therefore the limiting physical power flow solution, together with the corresponding augmented variables, is feasible for AR-OPF, A-OPF, and O-OPF and has the same nodal absorption $s$ as the initial point. This proves Item~1).
\end{IEEEproof}

\begin{IEEEproof}[Proof of Theorem~\ref{thm-revised-exactness}, Item~2)]
We use similar contradiction logic as in \cite{nickExactConvexFormulation2018}. Let $\omega^{\rm opt}$ be an optimal solution of AR-OPF. Assume, by contradiction, that it is inexact, i.e., that the relaxed \eqref{eq-AC-pf-f-soc} inequality is strict on at least one line. Then $d=    f^{(0)}-f^{(1)}    \neq0$. Applying \textit{Algorithm R.1} to $\omega^{\rm opt}$ and using Item~1 yields a recovered physical point $\omega^{\rm rec}$ that is feasible for AR-OPF, A-OPF, and O-OPF and satisfies $s^{\rm rec}=s^{\rm opt}$. By \eqref{eq-C4-prime} and Lemma~\ref{lem-revised-RIII}, $P_{1,\rm rec}^t < P_{1,\rm opt}^t$.

For the original AR-OPF objective,
\begin{equation}
J(\omega)
=
\sum_{l\in\mathcal L}
\mathcal C(p_l,q_l)
+
\mathcal C^e(P_1^t),
\end{equation}
and $s^{\rm rec}=s^{\rm opt}$ implies
\begin{equation}
\sum_{l\in\mathcal L}
\mathcal C(
p_l^{\rm rec},q_l^{\rm rec})
=
\sum_{l\in\mathcal L}
\mathcal C(
p_l^{\rm opt},q_l^{\rm opt}).
\end{equation}
Since $\mathcal C^e$ is strictly increasing, $P_{1,\rm rec}^t < P_{1,\rm opt}^t$ gives
\begin{equation}
    \mathcal C^e(P_{1,\rm rec}^t)
    <
    \mathcal C^e(P_{1,\rm opt}^t),
\end{equation}
and therefore $J(\omega^{\rm rec}) < J(\omega^{\rm opt})$.
This contradicts the optimality of $\omega^{\rm opt}$, since $\omega^{\rm rec}$ is also AR-OPF feasible.

Hence the assumed optimal solution cannot be inexact, and
\begin{equation}
f_l^{\rm opt}
=
\frac{
(P_{l,\rm opt}^t)^2
+
(Q_{l,\rm opt}^c)^2
}{
v_{\operatorname{up}(l),\rm opt}
},
\qquad
\forall l\in\mathcal L.
\end{equation}
Thus every optimal AR-OPF solution satisfies \eqref{eq-AC-pf-f} and is an optimal solution of A-OPF and O-OPF.
\end{IEEEproof}

\subsection{Impact of Slack Bus Reactive Power on Exactness}\label{sec-slack-Q-in-obj}

Above exactness result assumes that the cost associated with the external grid is a strictly increasing function of the active power imported from the slack bus. The argument can be extended to an objective depending on both active and reactive powers exchanged with the slack bus. Consider
\begin{equation}
    J_{\rm general} = \sum_{l \in \mathcal{L}}\left(\mathcal{C}\left(\mathfrak{R}(s_l), \mathfrak{I}(s_l) \right) \right) + \Psi(P_1^t,Q_1^t), \label{eq-AROPF-general-obj}
\end{equation}
where $\Psi:\mathbb R^2\rightarrow\mathbb R$ denotes the cost associated with the slack-bus power exchange. For fixed nodal injections $s$, the slack-bus active and reactive powers can be expressed as $P_1^t = \hat P_1^t+\mathbf a_P^\top f$ and $Q_1^t = \hat Q_1^t+\mathbf a_Q^\top f$. Let $\omega^{\mathrm{AR}}$ be an inexact feasible AR-OPF point and define $d:=f^{(0)}-f^{(1)}>0$. Under \eqref{eq-C2-prime}, the cumulative correction after the first iteration satisfies $ \left| \sum_{n=2}^{\infty}\Delta f^{(n)} \right| \le \mathbf R_Ed$. Consequently, for the recovered physical solution $\omega^{\mathrm{rec}}$,
\begin{equation}
\begin{aligned}
    P_{1,\mathrm{rec}}^t-P_{1,\mathrm{AR}}^t &\le -\mathbf a_P^\top d + \mathbf a_P^\top\mathbf R_Ed, \\
    Q_{1,\mathrm{rec}}^t-Q_{1,\mathrm{AR}}^t &\le -\mathbf a_Q^\top d + \mathbf a_Q^\top\mathbf R_Ed. 
\end{aligned}\label{eq-rec-AR-PQ}
\end{equation}
Therefore, if
\begin{equation}
    \mathbf a_P^\top\mathbf R_Ed < \mathbf a_P^\top d \quad \text{and} \quad \mathbf a_Q^\top\mathbf R_Ed < \mathbf a_Q^\top d, \label{eq-ap-aq-re}
\end{equation}
then $P_{1,\mathrm{rec}}^t<P_{1,\mathrm{AR}}^t$ and $Q_{1,\mathrm{rec}}^t<Q_{1,\mathrm{AR}}^t$.
This leads to the following extension of the exactness result.
\begin{corollary}
Suppose \eqref{eq-C1-prime}--\eqref{eq-C3-prime} hold and that the recovered power flow solution is feasible as established in the preceding theorem. Let the objective be given by \eqref{eq-AROPF-general-obj}, where $\Psi(P,Q)$ is component-wise non-decreasing in $P$ and $Q$. If, for every inexact optimal AR-OPF point, the physical-recovery procedure produces
\begin{equation}
\Psi(P_{1,\mathrm{rec}}^t,Q_{1,\mathrm{rec}}^t)<\Psi(P_{1,\mathrm{AR}}^t,Q_{1,\mathrm{AR}}^t), \label{eq-Psi-decrease}
\end{equation}
then the AR-OPF relaxation is exact.

In particular, condition \eqref{eq-ap-aq-re}, together with strict monotonicity of $\Psi$ in at least one of the corresponding arguments, are sufficient for \eqref{eq-Psi-decrease}.
\end{corollary}
\begin{IEEEproof}[Proof of Corollary 1]
    The recovered point $\omega^{\mathrm{rec}}$ has the same nodal power injections as the assumed optimal relaxed point $\omega^{\mathrm{AR}}$. Hence,
    $$
    \sum_{l \in \mathcal{L}}     \mathcal C(p_l^{\mathrm{rec}},q_l^{\mathrm{rec}}) = \sum_{l \in \mathcal{L}} \mathcal C(p_l^{\mathrm{AR}}, q_l^{\mathrm{AR}}).
    $$
    If \eqref{eq-Psi-decrease} holds, then $J(\omega^{\mathrm{rec}}) < J(\omega^{\mathrm{AR}})$, which contradicts the optimality of the inexact point $\omega^{\mathrm{AR}}$. Therefore every optimal AR-OPF solution must satisfy \eqref{eq-AC-pf-f} on all lines.
\end{IEEEproof}

\subsubsection{Linear active-reactive power objective}
An important special case is
\begin{equation}
    J_{\rm linear} = \sum_l\mathcal C(p_l,q_l) + \alpha P_1^t + \beta Q_1^t, \label{eq-AROPF-linear-obj}
\end{equation}
where $\alpha$ and $\beta$ are constant coefficients. Combining \eqref{eq-rec-AR-PQ} gives
$$
J(\omega^{\mathrm{rec}}) - J(\omega^{\mathrm{AR}}) \le{} - \left( \alpha\mathbf a_P^\top + \beta\mathbf a_Q^\top \right)d + \left(\alpha\mathbf a_P^\top + \beta\mathbf a_Q^\top \right) \mathbf R_Ed,
$$
provided $\alpha,\beta\ge0$. Hence the following condition directly guarantees exactness:
\begin{equation}
\left(\alpha\mathbf a_P^\top + \beta\mathbf a_Q^\top\right)\mathbf R_Ed < \left( \alpha\mathbf a_P^\top + \beta\mathbf a_Q^\top \right)d.\label{eq-exactness-PQ-RE-d}
\end{equation}
This gives the following special case of \textit{Corollary 1}.
\begin{corollary}\label{corollary-linear-obj-C4}
    For the linear slack-bus objective \eqref{eq-AROPF-linear-obj} with $\alpha\ge0,  \beta\ge0, (\alpha,\beta)\neq(0,0)$, condition \eqref{eq-exactness-PQ-RE-d}, together with \eqref{eq-C1-prime}--\eqref{eq-C3-prime}, is sufficient for exactness of AR-OPF. A solution-independent version of \eqref{eq-exactness-PQ-RE-d} is $\left( \alpha\mathbf a_P^\top + \beta\mathbf a_Q^\top \right) \mathbf R_E < \left( \alpha\mathbf a_P^\top + \beta\mathbf a_Q^\top \right)$.
\end{corollary}
Note that The original AR-OPF objective corresponds to $\beta=0$, in which case \eqref{eq-exactness-PQ-RE-d} reduces to \eqref{eq-C4-prime}.

\section{Explicit FR Construction}
\subsection{Directional Boundary Tracing via Losses-regularized Objective Function} \label{sec-directional-boundary-tracing-via-losses-obj}
\noindent{}Consider the losses-regularized directional objective function
$$
\begin{aligned}
J_{\rm \varepsilon} & = \cos\gamma \cdot P_1^t + \sin\gamma \cdot Q_1^t + \varepsilon\cdot (P_{\rm loss} + Q_{\rm loss})\\
& = (\varepsilon+\cos\gamma)\cdot P_1^t + (\varepsilon+\sin\gamma)\cdot Q_1^t - \varepsilon\cdot\sum_l (p_l+q_l).
\end{aligned}
$$
For the exactness argument, if $\varepsilon\ge1$, the relevant slack-bus coefficients
$\alpha_{\varepsilon,\gamma} = \varepsilon+\cos\gamma\ge0$, $\beta_{\varepsilon,\gamma} = \varepsilon+\sin\gamma\ge0$, and they cannot vanish simultaneously for any $\gamma$. Define $\mathbf c_{\varepsilon,\gamma}^{\top} := (\varepsilon+\cos\gamma)\mathbf a_P^\top+(\varepsilon+\sin\gamma)\mathbf a_Q^\top$.
According to \Cref{corollary-linear-obj-C4}, the solution-independent objective condition for exactness is
\begin{equation}
\mathbf c_{\varepsilon,\gamma}^{\top}\mathbf R_E < \mathbf c_{\varepsilon,\gamma}^{\top},
    \qquad
    \forall\gamma\in[0,2\pi).\label{eq-c-less-RE}
\end{equation}

\begin{corollary}\label{corollary-angle-independent}
Let $\mathbf B:=\mathbf I-\mathbf R_E$, $\mathbf u^\top:=\mathbf a_P^\top\mathbf B$, $\mathbf t^\top:=\mathbf a_Q^\top\mathbf B$. For a given $\varepsilon\ge1$, \eqref{eq-c-less-RE} is equivalent to:
\begin{equation}
    \varepsilon(\mathbf u+\mathbf t) > \sqrt{\mathbf u\circ\mathbf u+\mathbf t\circ\mathbf t}, \tag{C4'-FR}
    \label{eq-C4-prime-FR}
\end{equation}
where $\circ$ and the square root are interpreted component-wise.
\end{corollary}

\begin{IEEEproof}
Condition \eqref{eq-c-less-RE} is equivalent to
$$
    \varepsilon(\mathbf u^\top+\mathbf t^\top)
    +\cos\gamma\,\mathbf u^\top
    +\sin\gamma\,\mathbf t^\top>0 .
$$
For each $j$, $\min_{\gamma}\left(u_j\cos\gamma+t_j\sin\gamma\right)=-\sqrt{u_j^2+t_j^2}$. Therefore the inequality holds for every $\gamma$ if and only if $\varepsilon(u_j+t_j) > \sqrt{u_j^2+t_j^2}, \ \forall j$, which gives \eqref{eq-C4-prime-FR}.
\end{IEEEproof}

\begin{proposition}\label{theorem-monotonicity-epsilon}
Suppose \eqref{eq-C1-prime}--\eqref{eq-C3-prime} hold and $\varepsilon_1\ge1$. If \eqref{eq-c-less-RE} holds for every direction at $\varepsilon_1$, then it also holds for every $\varepsilon_2>\varepsilon_1$.
\end{proposition}

\begin{IEEEproof}
By \Cref{corollary-angle-independent}, exactness for all directions at $\varepsilon_1$ implies, component-wise, $\varepsilon_1(u_j+t_j)>\sqrt{u_j^2+t_j^2}$, and hence $u_j+t_j>0$. Therefore, for any
$\varepsilon_2>\varepsilon_1$,
$$
    \varepsilon_2(u_j+t_j)
    >
    \varepsilon_1(u_j+t_j)
    >
    \sqrt{u_j^2+t_j^2},
$$
for every $j$. Applying \Cref{corollary-angle-independent} again proves
the result.
\end{IEEEproof}

Consider the losses-regularized directional boundary tracing problem in \eqref{eq-opt-AR-OPF-epsilon},
\begin{equation}
    \begin{aligned}\label{eq-opt-AR-OPF-epsilon}
    \min_{s, S, v, f, \hat{S}, \bar{v}, \bar{S}, \bar{f}}&
    \quad 
    J_{\rm \varepsilon}
    \\
    \text{s.t.} &\quad \eqref{eq-AC-pf-St},  \eqref{eq-AC-pf-v},  \eqref{eq-AC-pf-Sb}, \eqref{eq-g-cons-res-P}, \eqref{eq-g-cons-res-Q}, \eqref{eq-AC-pf-f-soc}, \ \text{and}\\
    &\text{augmented con. (R.11), (R.12.c-R.12.g)},
\end{aligned}
\end{equation}
for a fixed regularization coefficient $\varepsilon$, define the continuously traced interface curve
$\Gamma_\varepsilon:= \left\{ y_{\rm AR,\varepsilon}^{\star}(\gamma):\gamma\in[0,2\pi)\right\}$. The finite set $\mathcal S_{\rm AR,\varepsilon}$ obtained from a discrete set of directions is regarded as a numerical approximation of $\Gamma_\varepsilon$. Let $\mathcal K_{\bar\varepsilon}$ denote the bounded region enclosed by $\Gamma_{\bar\varepsilon}$, i.e., all points on $\Gamma_{\bar\varepsilon}$ and the interior of $\Gamma_{\bar\varepsilon}$ that has a non-zero winding number, $\mathcal X_{\rm AR}$ denote the AR-OPF feasible set, and similarly $\mathcal X_{\rm AC}$ the O-OPF feasible set.

\begin{lemma}\label{lemma-compact-sets}
    The feasible sets $\mathcal X_{\rm AR}$ and $\mathcal X_{\rm AC}$, and the AC flexibility region $\mathcal Y_{\rm AC}$, are compact sets.
\end{lemma}
\begin{IEEEproof}
    Under finite nodal-injection, voltage, and ampacity limits, with $v^{\min}>0$, the physical variables of $\mathcal X_{\rm AC}$ are bounded. Moreover, the power flow equations and operational constraints define closed sets since their mappings are continuous on $v$. Hence $\mathcal X_{\rm AC}$ is closed and bounded, and therefore compact in finite dimensions \cite[Th.~2.41]{rudin1976principles}. The same argument applies to $\mathcal X_{\rm AR}$, whose auxiliary variables are bounded through the augmented and finite constraints. Finally, $\mathcal Y_{\rm AC}$ is the continuous coordinate projection of $\mathcal X_{\rm AC}$ and is therefore compact, and thus closed \cite[Th.~26.5]{munkres2000topology}.
\end{IEEEproof}



\begin{lemma}\label{lemma-continuity}
Under \eqref{eq-C1-prime}--\eqref{eq-C3-prime} and \eqref{eq-C4-prime-FR}, the optimal interface point admits a continuous selection $(\gamma,\varepsilon)\mapsto y_{\rm AR,\varepsilon}^\star(\gamma)$, and $y_{\rm AR,\varepsilon}^\star(\gamma)$ is unique.
\end{lemma}

\begin{IEEEproof}
Consider any two optimal AR-OPF solutions $\omega^a,\omega^b \in \mathcal{X}_{\rm AR}$ exist for a fixed $(\gamma,\varepsilon)$. Since $\mathcal{X}_{\rm AR}$ is convex and $J_\varepsilon$ is affine in $\omega$ by definition, $\omega^m=(\omega^a+\omega^b)/2$ is also optimal. All three solutions are also exact by \eqref{eq-C4-prime-FR}. For the branch that connected to the slack bus, define $\zeta=(P_1^t,Q_1^c)^\top$, \eqref{eq-f-convergence} and fixed $v_0$ give $v_0 f_1^i=\|\zeta^i\|^2, \ i\in\{a,b,m\}$. Since $f_1^m=(f_1^a+f_1^b)/2$ and $\zeta^m=(\zeta^a+
\zeta^b)/2$, the parallelogram identity yields
$$
0= \frac{\|\zeta^a\|^2+ \|\zeta^b\|^2}{2}-\left\|\frac{\zeta^a+\zeta^b}{2}\right\|^2=\frac14\|\zeta^a-\zeta^b\|^2.
$$
Thus $\zeta^a=\zeta^b$ and yields unique $y_{\rm AR,\varepsilon}^\star(\gamma)$ since $Q_1^t=Q_1^c-b_1v_0$, proving uniqueness of the optimal interface projection. Finally, recall that $\mathcal{X}_{\rm AR}$ is non-empty and compact, and $J_\varepsilon$ is jointly continuous in $(\omega,\gamma,\varepsilon)$. Berge's maximum theorem \cite[Th.~17.31]{aliprantis2006infinite} implies $(\gamma,\varepsilon)\mapsto y_{\rm AR,\varepsilon}^\star(\gamma)$ is continuous since the interface projection is unique.
\end{IEEEproof}

\begin{proposition}[Losses-dominated contraction]\label{proposition-losses-dominated-contraction}
 Let $L:=P_{\rm loss}+Q_{\rm loss}$ and $L^\star:=\min_{x\in\mathcal X_{\rm AR}}L(x)$, \Cref{lemma-continuity} indicates that the minimum-losses set has the singleton interface projection
    \begin{equation}
        \left\{
        y(x):x\in\arg\min_{\mathcal X_{\rm AR}}L(x)
        \right\}
        =\{y^\infty\}.
        \label{eq-unique-losses-projection}
    \end{equation}
    Then \begin{equation}
        \lim_{\varepsilon\rightarrow\infty}
        \sup_{\gamma\in[0,2\pi)}
        \left\|
        y_{\rm AR,\varepsilon}^{\star}(\gamma)-y^\infty
        \right\|=0.
        \label{eq-int-ass-infty}
    \end{equation}
\end{proposition}

\begin{IEEEproof}
By compactness of $\mathcal X_{\rm AR}$, let
$M:=\max_{x\in\mathcal X_{\rm AR}}\left\|y(x)\right\|<\infty$. For any minimum-losses solution $x^L$ and any direction $\gamma$, optimality of
$x_{\varepsilon,\gamma}^{\star}$ gives
$$
    L^\star
    \le
    L(x_{\varepsilon,\gamma}^{\star})
    \le
    L^\star+{2M}/{\varepsilon}.
$$
Hence $\sup_{\gamma}\left(    L(x_{\varepsilon,\gamma}^{\star})-L^\star    \right)   \rightarrow0$. 
Suppose \eqref{eq-int-ass-infty} were false. Then there would exist
$\delta>0$, a sequence $\varepsilon_k\rightarrow\infty$, and directions
$\gamma_k$ such that $\left\|y_{\rm AR,\varepsilon_k}^{\star}(\gamma_k) - y^\infty \right\| \ge\delta$. Compactness of $\mathcal X_{\rm AR}$ yields a convergent subsequence
$x_{\varepsilon_k,\gamma_k}^{\star}\rightarrow\bar x$.
By continuity of $L$ and the preceding losses bound, $ L(\bar x)=L^\star$, whereas continuity of $y(\cdot)$ gives $\|y(\bar x)-y^\infty\|\ge\delta$. This contradicts the unique minimum-losses interface projection \eqref{eq-unique-losses-projection}. Therefore \eqref{eq-int-ass-infty} holds.
\end{IEEEproof}

\begin{proposition}
\label{prop-interior-feasibility}
Suppose $\bar{\varepsilon}\ge1$, \eqref{eq-C1-prime}--\eqref{eq-C3-prime} and \eqref{eq-C4-prime-FR} hold at $\bar{\varepsilon}$ for every $\gamma\in[0,2\pi)$, then $\mathcal K_{\bar{\varepsilon}}\subseteq\mathcal Y_{\rm AC}$.
\end{proposition}

\begin{IEEEproof}
By \Cref{theorem-monotonicity-epsilon}, every directional AR-OPF optimum is exact for all $\varepsilon\ge\bar\varepsilon$, thus $\Gamma_\varepsilon\subseteq  \mathcal Y_{\rm AC},\ \forall\varepsilon\ge\bar\varepsilon$. Moreover, the established continuity and losses-dominated contraction results imply that $\Gamma_\varepsilon$ forms a continuous family converging uniformly to a point $y^\infty$, which is AC feasible by closedness of $\mathcal Y_{\rm AC}$. Consider any $y\in\mathcal K_{\bar\varepsilon}$. If $y\in\Gamma_{\bar\varepsilon}$ or $y=y^\infty$, its feasibility follows immediately. Otherwise, $\operatorname{wind}    (\Gamma_{\bar\varepsilon},y)\ne0$. Suppose no intermediate curve $\Gamma_\varepsilon$ passes through $y$. By continuity, the curves form a homotopy avoiding $y$, so their winding number around $y$ remains invariant. However, uniform convergence to $y^\infty\ne y$ implies that, for sufficiently large $\varepsilon$, the entire curve lies in a disk centered at $y^\infty$ that excludes $y$. Its winding number around $y$ must be zero, yielding a contradiction. Hence $y\in\Gamma_{\varepsilon_y}$ for some
$\varepsilon_y\ge\bar\varepsilon$, and is therefore AC feasible. This proves $\mathcal K_{\bar\varepsilon}\subseteq\mathcal Y_{\rm AC}$.
\end{IEEEproof}

\subsection{FR Approximation Algorithms}

The preceding analysis provides a constructive procedure for numerically obtaining an explicit inner approximation of the FR. Two variants are considered. The first targets the largest boundary obtained with the smallest $\varepsilon$ for which exactness can be certified ex-ante or numerically verified ex-post. The second additionally seeks a numerically convex representation by progressively increasing $\varepsilon$.

\subsubsection{Adaptive explicit FR approximation}
\label{sec-adaptive-FR}

For a given $\varepsilon$, let $\Gamma_N=\{\gamma_1,\ldots,\gamma_{N_\gamma}\}$ be a sufficiently dense set of tracing directions, and define the sampled boundary $\mathcal S_{\varepsilon} = \left\{ y_{\rm AR,\varepsilon}^{\star}(\gamma_i): \gamma_i\in\Gamma_N \right\}$. The points are ordered according to $\gamma_i$ and connected cyclically to form the polygonal boundary $\widehat{\Gamma}_{\varepsilon}$, whose enclosed region is $\widehat{\mathcal K}_{\varepsilon}$. For a specific $\varepsilon$, exactness is first checked using the proposed ex-ante sufficient conditions. When these conditions are not satisfied, exactness is instead assessed numerically at all sampled boundary points. Define the inexactness
\begin{equation}
\xi_{\rm ex}
= \max_{l} \left(f_l v_{{\rm up}(l)} - |S_l^t + j v_{{\rm up}(l)} b_l|^2\right), \label{eq-expost-exactness}
\end{equation}
and regard solutions as numerically exact if
$\xi_{\rm ex}\le\tau_{\rm ex}$.

\begin{algorithm}[ht]
\caption{Adaptive explicit FR approximation}
\label{alg-adaptive-inner}
\begin{algorithmic}[1]
\STATE \textbf{Input:}
$N_\gamma$, $\Delta\varepsilon$, $\tau_{\rm ex}$
\STATE Set $\varepsilon\gets1$
\WHILE{true}
    \STATE Evaluate \eqref{eq-C1-prime}--\eqref{eq-C3-prime}, \eqref{eq-C4-prime-FR}
    \STATE Solve \eqref{eq-opt-AR-OPF-epsilon}
    for all $\gamma_i\in\Gamma_N$, construct $\mathcal S_\varepsilon$, $\widehat{\Gamma}_\varepsilon$, and $\widehat{\mathcal K}_\varepsilon$

    \IF{\eqref{eq-C1-prime}--\eqref{eq-C3-prime} and \eqref{eq-C4-prime-FR} hold}
        \STATE \textbf{return}
        $\widehat{\mathcal K}_\varepsilon$
        as the discretization of $\mathcal K_\varepsilon$
    \ELSE
        \STATE Compute $\xi_{\rm ex}$ using
        \eqref{eq-expost-exactness}
        \IF{$\xi_{\rm ex}\le\tau_{\rm ex}$}
            \STATE \textbf{return}
            $\widehat{\mathcal K}_\varepsilon$
            as a numerically validated inner approximation
        \ENDIF
    \ENDIF

    \STATE $\varepsilon\gets
    \varepsilon+\Delta\varepsilon$
\ENDWHILE
\end{algorithmic}
\end{algorithm}

The adaptive procedure is summarized in Algorithm~\ref{alg-adaptive-inner}. When the ex-ante conditions hold, the continuous region
$\mathcal K_\varepsilon$ is certified to satisfy $\mathcal K_\varepsilon\subseteq\mathcal Y_{\rm AC}$ by Proposition~\ref{prop-interior-feasibility}. Otherwise, the ex-post test verifies exactness at the sampled directions, and $\widehat{\mathcal K}_\varepsilon$ is interpreted as a numerically validated approximation.

\subsubsection{Adaptive numerically convex approximation}
\label{sec-adaptive-convex}

A region obtained from Algorithm~\ref{alg-adaptive-inner} may remain non-convex. To obtain a numerically convex representation, the same adaptive search over $\varepsilon$ is supplemented by a direct comparison between the sampled region and its convex hull.

For a sampled region
$\widehat{\mathcal K}_\varepsilon$, define $\widehat{\mathcal H}_\varepsilon := \operatorname{conv} (\widehat{\mathcal K}_\varepsilon)$. Since  $\widehat{\mathcal K}_\varepsilon \subseteq \widehat{\mathcal H}_\varepsilon$, the sampled region is convex if and only if the reverse inclusion also holds. Numerically, we allow a small geometric tolerance
$\tau_{\rm cvx}$ and require $\widehat{\mathcal H}_\varepsilon\subseteq\widehat{\mathcal K}_\varepsilon \oplus B_{\tau_{\rm cvx}}$, where $B_{\tau_{\rm cvx}}$ is the closed disk of radius
$\tau_{\rm cvx}$ that accounting for numerical tolerance and $\oplus$ denotes the Minkowski sum. Equivalently, every point of the convex hull must lie within distance
$\tau_{\rm cvx}$ of the sampled region.

\begin{algorithm}[ht]
\caption{Adaptive numerically convex FR approximation}
\label{alg-adaptive-convex}
\begin{algorithmic}[1]
\STATE \textbf{Input:}
$N_\gamma$, $\Delta\varepsilon$,
$\tau_{\rm ex}$, $\tau_{\rm cvx}$
\STATE Set $\varepsilon\gets1$,
$\widehat{\mathcal K}_{\rm ref}\gets\varnothing$

\WHILE{true}
    \STATE Evaluate \eqref{eq-C1-prime}--\eqref{eq-C3-prime}, \eqref{eq-C4-prime-FR}
    \STATE Solve \eqref{eq-opt-AR-OPF-epsilon}
    for all $\gamma_i\in\Gamma_N$, construct $\mathcal S_\varepsilon$, $\widehat{\Gamma}_\varepsilon$, and $\widehat{\mathcal K}_\varepsilon$

    \IF{\eqref{eq-C1-prime}--\eqref{eq-C3-prime} and \eqref{eq-C4-prime-FR} hold}
        \STATE Set $\texttt{exact}\gets\texttt{true}$
    \ELSE
        \STATE Compute $\xi_{\rm ex}$ using
        \eqref{eq-expost-exactness}, set $\texttt{exact}\gets(\xi_{\rm ex}\le\tau_{\rm ex})$
    \ENDIF

    \IF{$\texttt{exact}$}
        \IF{$\widehat{\mathcal K}_{\rm ref}=\varnothing$}
            \STATE Save the first accepted inner region:
            $\widehat{\mathcal K}_{\rm ref}
            \gets
            \widehat{\mathcal K}_\varepsilon$
        \ENDIF

        \STATE Compute
        $\widehat{\mathcal H}_\varepsilon
        \gets
        \operatorname{conv}
        (\widehat{\mathcal K}_\varepsilon)$

        \IF{
        $\widehat{\mathcal H}_\varepsilon
        \subseteq
        \widehat{\mathcal K}_{\rm ref}
        \oplus B_{\tau_{\rm cvx}}$
        }
            \STATE \textbf{return}
            $\widehat{\mathcal H}_\varepsilon \ominus B_{\tau_{\rm cvx}}$
            as the numerically convex inner approximation
        \ENDIF
    \ENDIF

    \STATE $\varepsilon\gets
    \varepsilon+\Delta\varepsilon$
\ENDWHILE
\end{algorithmic}
\end{algorithm}


Algorithm~\ref{alg-adaptive-convex} first identifies an analytically certified or numerically validated inner region $\widehat{\mathcal K}_{\rm ref}$, which is retained as a reference feasible region. For subsequent values of $\varepsilon$, the convex hull $\widehat{\mathcal H}_\varepsilon= \operatorname{conv}(\widehat{\mathcal K}_\varepsilon)$ is constructed and tested against this reference region. The procedure terminates when $\widehat{\mathcal H}_\varepsilon\subseteq    \widehat{\mathcal K}_{\rm ref}\oplus B_{\tau_{\rm cvx}}$. To compensate for the geometric tolerance introduced by this test, we define the eroded hull $\widehat{\mathcal H}^{\,\rm safe}_\varepsilon:=
\widehat{\mathcal H}_\varepsilon \ominus B_{\tau_{\rm cvx}}$, where $\ominus$ denotes Minkowski erosion. The containment $\widehat{\mathcal H}^{\,\rm safe}_\varepsilon\subseteq\widehat{\mathcal K}_{\rm ref}$ is then verified numerically before acceptance. Since erosion of a convex set by a disk preserves convexity, the returned set $\widehat{\mathcal H}^{\,\rm safe}_\varepsilon$
provides a numerically convex inner approximation of the AC flexibility region.

When several branches are connected to the same slack bus, Algorithm~\ref{alg-adaptive-convex} can be applied separately to each sub-network to obtain $\widehat{\mathcal H}^{\,\rm safe}_i$. The aggregated FR is then obtained by their Minkowski sum,
$$
\widehat{\mathcal H}_{\rm agg}=\bigoplus_{i=1}^{N_{\rm sub}}\widehat{\mathcal H}^{\,\rm safe}_i.
$$

\section{Numerical Example}\label{sec-numerical-example}

\subsection{Test systems and implementation}

The proposed methods are evaluated using an illustrative 4-node network and a collection of larger radial networks to assess both their numerical behavior and scalability. The two algorithms described in the previous section are implemented in Julia programming language using JuMP~v1.30.1 \cite{Lubin2023}. The nonlinear O-OPF boundary tracing problems are solved with IPOPT \cite{wachterImplementationInteriorpointFilter2006}, while the convex AR-OPF problems are solved with Clarabel \cite{Clarabel_2024}. All simulations are performed on a laptop equipped with an Apple M5 Pro processor and 48~GB of RAM. Unless otherwise stated, the directional boundary is discretized using
$N_\gamma=50$ equally spaced angles. The ex-post exactness tolerance is set to
$\tau_{\rm ex}=5\times10^{-4}$, while the geometric tolerance used in
Algorithm~\ref{alg-adaptive-convex} is
$\tau_{\rm cvx}=5\times10^{-3}$. These values are kept unchanged throughout the numerical experiments unless explicitly specified otherwise.

\subsection{Illustrative 4-node network}

\begin{figure}[ht]
    \centering
    \includegraphics[width=.8\linewidth]{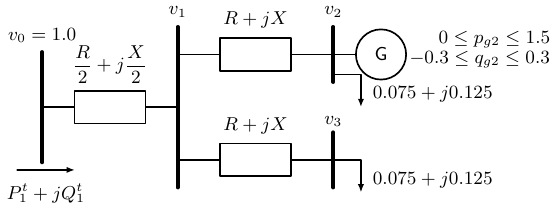}
    \caption{4-node test network, in which $R=0.035, X=0.07$.}
    \label{fig-4-node-topo}
\end{figure}
We first use the 4-node radial network shown in \Cref{fig-4-node-topo} to illustrate the behavior of the proposed boundary tracing and convexification procedures. The system is adapted from \cite{prionistisFlexibilityRegionDiscontinuity2025} and consists of 1 flexible generator connected at bus~2 and fixed loads at buses~2 and~3. The squared slack voltage is fixed at $v_0=1.0$, while the non-slack voltage limits are $v^{\max}=1.05^2$ and $v^{\min}=0.95^2$. The squared branch-current limits $I^{\max}=1.0^2$ are added to for all lines. The line parameters are $R=0.035$ and $X=0.07$, as indicated in \Cref{fig-4-node-topo}.
\begin{figure}[ht]
    \centering
    \includegraphics[width=\linewidth]{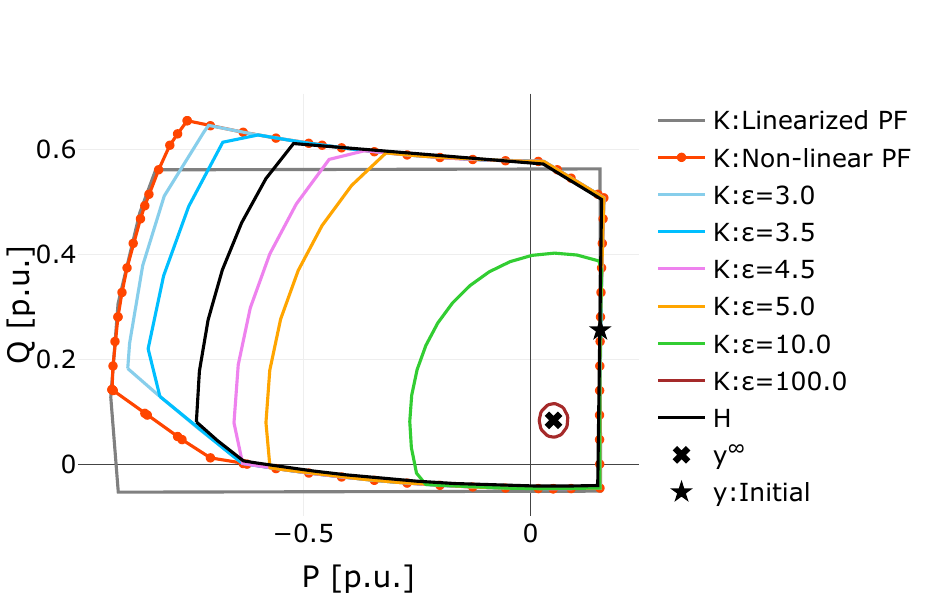}
    \caption{Boundary tracing results of the 4-node network. The initial operating point (star) corresponds to the baseline operation without adjustments of flexible resources.}
    \label{fig-FR-4-node}
\end{figure}

The boundary tracing results of the 4-node test network are shown in \Cref{fig-FR-4-node}. Several observations can be made. First, the proposed method is compared against two commonly used FR estimation baselines. The curve denoted by \emph{Linearized PF} is obtained from a fixed-point linearization of the power flow equations following \cite{bernsteinLinearPowerflowModels2017}. Such linearized models are widely adopted in multi-period and aggregated FR estimation because of their computational tractability \cite{cuiNetworkCognizantTimeCoupledAggregate2021,jacobsProbabilisticFlexibilityAggregation2026}. In the present example, however, the resulting region differs noticeably from that obtained with the full nonlinear AC model, particularly along the left and lower parts of the boundary. The \emph{Non-linear PF} curve is obtained by conventional directional scanning of the non-approximated AC power flow equations, following the boundary tracing approaches proposed in \cite{silvaEstimatingActiveReactive2018,capitanescuTSODSOInteraction2018}, and is therefore used as the numerical reference. Note that \emph{Non-linear PF} curve does not provide any guarantee of operational feasibility within its boundary. A slight non-convexity can be observed along the upper boundary, illustrating that the AC flexibility region is not necessarily convex even for this small radial network, as discussed in \cite{pozoTSODSOFlexibilityRegions2025}. The boundaries generated by Algorithm~1 for different values of $\varepsilon$ are also shown in \Cref{fig-FR-4-node}. For relatively small $\varepsilon$, the AR-OPF boundary closely follows the nonlinear AC reference, including its nonconvex features. As $\varepsilon$ increases, the boundary contracts progressively toward the minimum-losses point $y^\infty$, consistently with the losses-dominated behavior discussed in \Cref{sec-directional-boundary-tracing-via-losses-obj}. Algorithm~\ref{alg-adaptive-convex} exploits this contraction to construct the numerically convex inner set $\widehat{\mathcal H}^{\,\rm safe}$, while retaining the non-approximated AC network model and the proposed exactness verification framework. For this 4-node case, both the original conditions R.C1--R.C5 and the proposed conditions \eqref{eq-C1-prime}--\eqref{eq-C3-prime}+\eqref{eq-C4-prime-FR} are satisfied. The ex-post relaxation residuals of all tested AR-OPF solutions are smaller than the order of $10^{-8}$, indicating that the relaxation is numerically exact for this example.

\subsection{Scalability and conservativeness}
\label{sec-numerical-scalability}

To evaluate the scalability and conservativeness of the proposed methods, we consider seven radial networks ranging from 12 to 141 nodes, adapted from MATPOWER test cases \cite{zimmerman_2025_15871662}, together with a synthetic Swiss distribution network (Swiss-PDG) developed in \cite{guptaCountrywidePVHosting2021,onetoLargescaleGenerationGeoreferenced2025}. The latter corresponds to the network originally indexed as \texttt{1024-10\_0\_6} with 87 nodes and includes shunt components. Since the selected MATPOWER cases do not originally include distributed flexible resources, flexible generators and storage systems with box-constrained power injections are added at randomly selected non-slack buses, following the configuration of the previous 4-node network. In some networks, flexible loads capable of reducing their power by up to 10\% are also included. For the sake of reproducibility, the complete test datasets are publicly available.\footnote{\url{https://github.com/DESL-EPFL/AROPF-FR-approximation}} For all cases, $v_0=1.0$ is fixed, while the non-slack voltage limits are set to $v^{\min}=0.9^2$ and $v^{\max}=1.1^2$. Branch current limits are specified according to the scale of the corresponding networks. The numbers of radial sub-networks connected to the slack bus, $N_{\rm sub}$, and flexible resources, $N_{\rm flex}$, are summarized in \Cref{tab:comparison}.

\begin{table*}[t]
\caption{Comparison of exactness conditions, approximation
accuracy, and computational performance across test networks.}
\label{tab:comparison}
\centering
\begin{tabular}{ccccccccccccc}
\toprule
Network & $N_{\mathrm{sub}}$ & $N_{\mathrm{flex}}$ & R.Cx & Cx' & $\xi_{\rm ex}^{\rm max}$ & Area ratio & $t^{\mathrm{NLP}}_{\mathrm{Ipopt}}$ & $t_{\mathrm{Clarabel}}^{\rm AROPF}$ & $t_{\mathrm{Ipopt}}^{\rm AROPF}$ & $M^{\mathrm{NLP}}_{\mathrm{Ipopt}}$ & $M_{\mathrm{Clarabel}}^{\rm AROPF}$ & $M_{\mathrm{Ipopt}}^{\rm AROPF}$ \\
& & & & & & & [s] & [s] & [s] & [MB] & [MB] & [MB]\\
\midrule
4-node & 1 & 1 & $\checkmark$ & $\checkmark$ & \num{2e-8} & 0.74 & 0.061 & 0.357 & 0.461 & 6.471 & 284.5 & 20.45 \\
12-node & 1 & 13 & $\times$ & $\checkmark$ & \num{1e-8} & 0.79 & 0.110 & 0.211 & 0.385 & 17.75 & 188.1 & 12.48 \\
33-node & 1 & 10 & $\times$ & $\checkmark$ & \num{3e-9} & 0.95 & 0.205 & 1.006 & 1.154 & 55.34 & 711.3 & 31.40 \\
74-node & 3 & 42 & $\times$ & $\checkmark$ & \num{2e-5} & 0.95 & 0.374 & 1.491 & 2.648 & 161.6 & 1388 & 74.82 \\
85-node & 1 & 42 & $\times$ & $\checkmark$ & \num{3e-6} & 0.97 & 0.448 & 2.131 & 2.972 & 191.7 & 1816 & 73.37 \\
94-node & 1 & 68 & $\times$ & $\times$ & \num{2e-6} & 0.99 & 0.565 & 2.783 & 3.722 & 231.1 & 2270 & 99.67 \\
118-node & 3 & 21 & $\times$ & $\checkmark$ & \num{3e-8} & 0.98 & 0.663 & 3.422 & 4.692 & 324.7 & 2820 & 108.9 \\
141-node & 1 & 67 & $\times$ & $\times$ & \num{1e-6} & 0.46 & 0.898 & 10.09 & 13.87 & 446.3 & 7808 & 254.3 \\
Swiss-PDG & 4 & 41 & $\times$ & $\checkmark$ & \num{4e-7} & 0.93 & 0.514 & 2.861 & 3.145 & 198.4 & 2233 & 118.8 \\
\bottomrule
\end{tabular}
\end{table*}

 \begin{figure*}[t]
     \centering     \includegraphics[width=\linewidth]{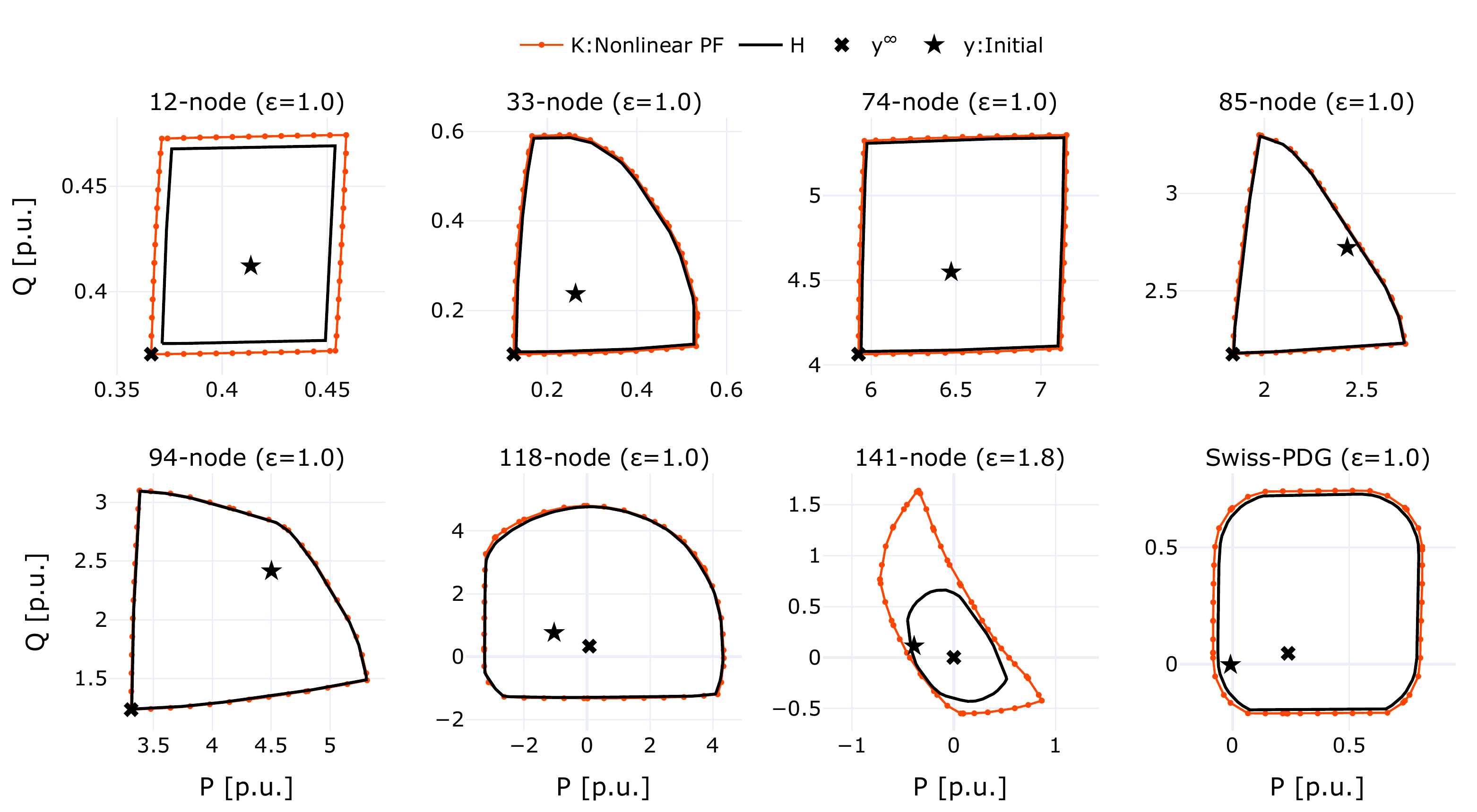}
     \caption{Comparison of FR
boundaries given by nonlinear power flow and the numerically convex inner approximations $\mathcal H$
obtained by Algorithm~\ref{alg-adaptive-convex} for different
distribution networks. The initial operating point (star) corresponds to the baseline operation without adjustments of flexible resources.}
\label{fig:boundaries}
 \end{figure*}

\Cref{fig:boundaries} compares $\widehat{\mathcal H}^{\,\rm safe}$ constructed by Algorithm~\ref{alg-adaptive-convex} against the FR boundaries obtained from nonlinear O-OPF scanning. The minimum-losses points $y^\infty$ are also indicated. For most networks, the proposed convex approximations closely reproduce the nonlinear boundaries, despite variations in network size, topology, and the number of flexible resources. Particularly close agreement is observed for the 33-, 74-, 85-, 94-, and 118-node networks, where the retained FR exceeds 95\% of the nonlinear reference. The Swiss-PDG case also achieves an area ratio of 0.93, demonstrating the applicability of the method to a network with shunt components and multiple radial sub-networks. The 141-node network presents a more challenging case. Its nonlinear FR exhibits pronounced non-convexity, and the proposed approximation retains only 46\% of the reference area. This illustrates the inherent trade-off between convexity and conservativeness: although the losses-regularized formulation enables the construction of a numerically feasible convex region, some FR may be excluded when the original region is strongly nonconvex. Therefore, the proposed method does not guarantee uniformly low conservativeness across different network configurations.

The quantitative results are summarized in \Cref{tab:comparison}, where R.Cx denotes R.C1--R.C5, while Cx' denotes \eqref{eq-C1-prime}--\eqref{eq-C3-prime}+\eqref{eq-C4-prime-FR}. The original conditions R.C1--R.C5 are satisfied only for the 4-node example, whereas the revised conditions \eqref{eq-C1-prime}--\eqref{eq-C3-prime}+\eqref{eq-C4-prime-FR} are satisfied for seven of the nine networks. This demonstrates the reduced conservativeness of the proposed sufficient conditions. For the 94- and 141-node cases, neither set of sufficient conditions is fulfilled; nevertheless, the ex-post residuals $\xi_{\rm ex}$ remain far below the prescribed tolerance, supporting the numerical exactness of the computed AR-OPF solutions even when the analytical certificates are inconclusive. \Cref{tab:comparison} also reports the total computation time $t$ and memory consumption $M$ of the nonlinear O-OPF and AR-OPF based boundary tracing, in seconds and MB, respectively. Three solver configurations are considered: nonlinear PF solved by Ipopt, AR-OPF solved by Clarabel, and AR-OPF solved by Ipopt. Clarabel is employed as a dedicated conic optimizer that exploits the convex formulation, while Ipopt is additionally applied to AR-OPF to provide a controlled comparison under the same general-purpose nonlinear solver used for the O-OPF baseline. This distinction helps separate formulation-dependent computational cost from solver-dependent performance. The nonlinear PF formulation consistently exhibits the lowest computation time. For AR-OPF, Clarabel generally outperforms Ipopt in computation time, although both require more time than the nonlinear baseline. The computational complexity is attributable in part to the additional auxiliary variables and augmented constraints introduced by AR-OPF. In terms of memory consumption, AR-OPF solved by Ipopt requires less memory than nonlinear PF solved by the same solver in most cases.

\section{Discussion}

The numerical results in \Cref{sec-numerical-example} demonstrate that the proposed methods can construct explicit convex inner approximations for networks of different sizes and configurations, with high area retention and numerically exact SOCP relaxations in most tested cases. The revised, less-conservative sufficient conditions enlarge the range of operating conditions for which exactness can be certified ex ante. Unlike conventional nonlinear boundary tracing, which generally provides locally optimal boundary samples without a global interior-feasibility guarantee, the proposed framework can certify the AC feasibility of the entire enclosed region under the established sufficient conditions. The resulting explicit convex representations are particularly relevant to TSO-DSO coordination, as they can be directly embedded into higher-level optimization problems without introducing additional nonconvex feasibility constraints. Moreover, the inner approximation property ensures that the flexibility requested by the TSO remains within the admissible operating region of the underlying distribution network, subject to the stated numerical accuracy. Although the proposed approach trades a slightly larger computational cost to obtain a numerically convex FR set, it is capable to assess the FR in a matter of seconds, guaranteeing its use in actual operation context of TSO-DSO coordination.  Compared with related SOCP-based approaches \cite{prionistisFastMethodApproximate2023,moringExploringInexactnessSecondOrder2026}, no significant relaxation inexactness is observed in the tested cases. Nevertheless, computational performance, the conservativeness of the sufficient conditions, and possible flexibility losses in strongly nonconvex regions remain limitations.

\section{Conclusion}

This paper proposed a framework for constructing explicit inner approximations of TSO-DSO FRs based on exact convex relaxations of the optimal power flow in radial  networks. First, less-conservative sufficient conditions for AR-OPF exactness were derived, extending the existing results and improving the applicability of ex-ante certification. Furthermore, sufficient conditions were established under which the entire region enclosed by the continuously traced boundary is AC feasible. Building on these theoretical results, two adaptive algorithms were developed to construct explicit inner FRs and numerically convex representations. Numerical evaluations on radial networks of different scales demonstrated consistently small relaxation residuals and high flexibility-area retention in most cases. The proposed framework therefore provides a theoretically grounded approach to obtaining operationally admissible flexibility representations for TSO-DSO coordination thanks to a computationally tractable algorithm. Future work will focus on: 1) reducing the computational burden and conservativeness of the approximations, particularly for strongly nonconvex FRs; 2) extending the approach to meshed networks; 3) integrating time-coupling resource constraints and cost curves.

\bibliographystyle{IEEEtran}
\bibliography{AROPF-FR-Paper}


\end{document}